\documentclass[showpacs,amsmath,amsfont,amssymb,pre,aps,superscriptaddress,reprint]{revtex4-1}
\usepackage{graphicx}
\usepackage{color}
\usepackage{bbm}
\usepackage{CJK}
\begin{document}

\title{Validity of Fourier’s law in one-dimensional momentum-conserving lattices with asymmetric interparticle interactions}

\author{Lei Wang}
\affiliation{Department of Physics, Renmin University of China, Beijing 100872, People's Republic of China}
\affiliation{Department of Physics and Centre for Computational Science and Engineering, National University of Singapore, Singapore 117546}
\author{Bambi Hu}
\affiliation{Department of Physics, University of Houston, Houston, Texas 77204-5005, USA}
\affiliation{Department of Physics and Centre for Computational Science and Engineering, National University of Singapore, Singapore 117546}
\author{Baowen Li}
\affiliation{Center for Phononics and Thermal Energy Science,
             School of Physical Science and Engineering, Tongji University, 200092, Shanghai, People' s Republic of China}
\affiliation{Department of Physics and Centre for Computational Science and Engineering, National University of Singapore, Singapore 117546}

\date{\today}

\begin{abstract}
We have numerically studied heat conduction in a few one-dimensional momentum-conserving lattices with asymmetric interparticle interactions by the nonequilibrium heat bath method, the equilibrium Green-Kubo method, and the heat current power spectra analysis. Very strong finite-size effects are clearly observed. Such effects make the heat conduction obey a Fourier-like law in a wide range of lattice lengths.
However, in yet longer lattice lengths, the heat conductivity regains its power-law divergence.
Therefore the power-law divergence of the heat conductivity in the thermodynamic limit is verified,
as is expected by many existing theories.
\end{abstract}

\maketitle

\section{introduction}

Heat conduction induced by a small temperature gradient in the stationary state
is expected to satisfy Fourier's law: $j=-\kappa \nabla T$.
A complete understanding of the microscopic mechanism that determines
this law remains a challenging problem of
nonequilibrium statistical mechanics~\cite{PhysRep.377.1,*AdvPhys.57.457}.
Fourier's law implies that a fixed temperature difference $\Delta T$ that is applied to a homogeneous material with length $L$ induces a
stationary state heat current $j$ that should be inversely proportional to $L$:
$j=-\kappa \Delta T/L$.
On the other hand, $j$ is found, by numerical simulation, to decay as $L^{-1+\alpha}$
with a positive $\alpha$ in many one-dimensional (1D) lattice models~\cite{PhysRevLett.78.1896,*PhysRevE.59.R1,*PhysRevLett.98.184301}.
This implies an infinite $\kappa$ which diverges with $L$ as $L^{\alpha}$ in the
thermodynamic limit.
When an on-site potential is absent, i.e., the particles in the lattice only interact with other particles,
the total momentum is conserved.
It has been generally accepted for many years, with only very few exceptions such as the coupled rotator lattice~\cite{PhysRevLett.84.2144,*PhysRevLett.84.2381},
that heat conduction in such a lattice belongs to this class~\cite{PhysRep.377.1,*AdvPhys.57.457,PhysRevLett.84.2857}.
Many theories, e.g., renormalization group~\cite{PhysRevLett.89.200601,*PhysRevE.73.061202}
and mode coupling~\cite{PhysRevE.73.060201,EurophysLett.43.271,*PhysRevE.68.067102}, support such a conclusion.
As for the detailed value of the divergency exponent $\alpha$, however, different theories suggest differently.
Mainstream expectations include $1/3$~\cite{PhysRevLett.89.200601,*PhysRevE.73.061202,PhysRevE.73.060201}, $2/5$~\cite{EurophysLett.43.271,*PhysRevE.68.067102},
and $1/2$~\cite{PhysRevLett.96.204303,*PhysRevLett.108.180601}. More details can also be found in the review articles Ref.~\cite{PhysRep.377.1,*AdvPhys.57.457}.

It is worth mentioning that such a topic is not merely a purely academic issue.
The rapid progress nowadays in nanotechnology has already enabled us to experimentally measure the size dependence of the heat conductivity
in many one-dimensional~\cite{PhysRevLett.101.075903} and two-dimensional~\cite{NatMater.9.555,*NanoLett.11.113,*NanoLett.12.3238} microscopic materials.
The topic thus also has extensive application values~\cite{RevModPhys.84.1045}.

\section{model and simulation}

Among various models, the lattice model, due to its simplicity, has been studied the most.
The Hamiltonian of a 1D lattice takes the form
\begin{equation}
H =\sum_i \frac{\dot x^2_i}2+V(x_i-x_{i+1}),
\label{eq:Ham}
\end{equation}
where $x_i$ is the displacement of the $i$th particle relative to its equilibrium position.
Without an on-site potential, the lattice is total momentum conserving.
In this paper we focus on the lattices with asymmetric interparticle interactions, namely $V(x)\neq V(-x)$.

The first one we study is the Fermi-Pasta-Ulam (FPU)-$\alpha\beta$ lattice,
\begin{equation}
 V(x)=\frac{1}{2}k_2 x^2+\frac{1}{3} k_3 x^3+\frac{1}{4}k_4 x^4.
\label{eq:fpuab}
\end{equation}
Throughout the paper we set $k_2=1$ ,$k_3=2$, and $k_4=1$.
This is the simplest model with asymmetric interparticle interactions.
We study its heat conduction by the nonequilibrium heat bath method first.
Fixed boundary conditions are applied to a lattice with length $L$. The left- and rightmost ends of the lattice are
coupled to Langevin heat baths with temperatures $T_L=1.5$ and $T_R=0.5$, respectively.
In the nonequilibrium stationary state, the heat conductivity is defined as:
\begin{equation}
 \kappa_{\rm NE}(L)\equiv\frac J{\nabla T},
\label{eq:kne}
\end{equation}
where $J$ is the heat current and $\nabla T$ is the temperature gradient in the lattice.
A number of independent
runs starting from different randomly chosen initial states are performed.
The simulation time depends on the lattice size. For the longest lattice ($L=65536$)
the average is performed over $4\times 10^8$ dimensionless time units.
A very flat heat conductivity is observed in the lattice length $L$ ranging from several hundred to ten thousand.
However, in yet longer lattices, say, $L>1\times10^4$, the running slope of the heat conductivity $\kappa$ grows again.
In fact such a curving up of $\kappa$ was reported in our previous work~\cite{EurophysLett.93.54002}.
But in this study the asymmetric term $k_3=2$ is much greater than that in the previous work.
By comparing the two results, we see that even the asymmetry is increased to so high a value
($k_3=2$ is the maximum value that keeps the potential single well, at the given $k_2=k_4=1$.),
the curving up of $\kappa$ can only be postponed,
but its asymptotic behavior is not affected.

\begin{figure}[ht]
\includegraphics[width=\columnwidth]{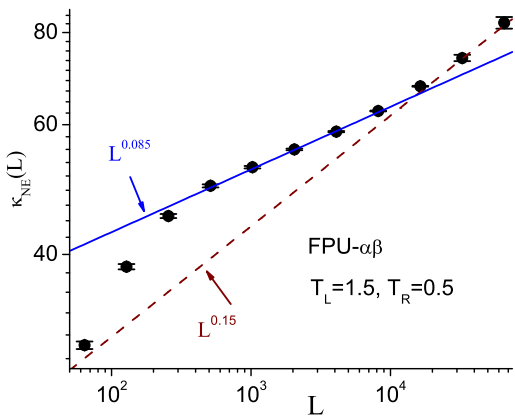}
\vspace{-1.0cm} \caption{\label{fig:fpua2T1heat} (color online).
The heat conductivity $\kappa_{\rm NE}$ as a function of the lattice length $L$.
The increase of $\kappa_{\rm NE}$ is slow in a regime of $L<1\times10^4$; however, it begins to speed up at about $L \sim 2 \times 10^4$.
Lines with different slopes are drawn for reference.}
\end{figure}

We next use the equilibrium Green-Kubo method~\cite{GREENKUBO},
which provides an alternative way to determining $\alpha$ with much higher cost efficiency~\cite{PhysRevLett.105.160601,EurophysLett.93.54002,PhysRevE.86.040101}.
The rescaled heat current autocorrelation function is defined as
\begin{equation}
c(\tau)\equiv\lim_{N\rightarrow\infty}\frac1{k_BT^2N}<J(t)J(t+\tau)>_t,
\label{eq:JJ}
\end{equation}
where $J(t)\equiv\sum_i j_i(t)$ is the instantaneous global heat
current and $N$ is the total number of particles. The Boltzmann constant $k_B$ is set to unity.
The simulations are carried out in lattices with periodic boundary conditions since they provide the best convergence to the thermodynamic limit.
Microcanonical simulations are applied with zero total
momentum and identical energy density $\epsilon=0.846$ that corresponds to the temperature $T=1$.

\begin{figure}[ht]
\includegraphics[width=\columnwidth]{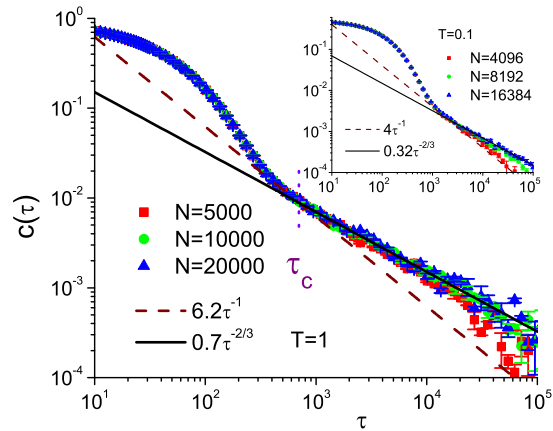}
\vspace{-1.0cm} \caption{\label{fig:JJfpuaT1} (color online).
The rescaled heat current autocorrelation function $c(\tau)$ of the FPU-$\alpha\beta$ lattice, for $T=1$ and various particle numbers $N$.
Lines with slope $-1$ and $-2/3$ are drawn for reference.
Inset: The same lattices but for temperature $T=0.1$.
This lower value of $T$ enlarges the threshold time lag $\tau_c$,
but it does not change
the asymptotic decay of $c(\tau)$, which remains slower than $\tau^{-1}$.
}
\end{figure}

Fig.~\ref{fig:JJfpuaT1} depicts $c(\tau)$ as a function of the time lag $\tau$ for various particle numbers $N$.
We see that in the short-$\tau$ regime, $c(\tau)$ follows a very fast exponential-like decay,
just like what was reported in Ref.~\cite{2012arXiv1204.5933C}.
However, when $\tau$ is greater than a threshold value $\tau_c=700$, $c(\tau)$ recovers a slow power-law decay.
A reasonable estimate of the asymptotic exponent for this certain model,
which is also supported by a theoretical expectation~\cite{PhysRevLett.89.200601,*PhysRevE.73.061202}, is $-2/3$.
In Ref.~\cite{2012arXiv1204.5933C} because $c(\tau)$ is studied in a short range of values of $\tau$ ($\tau<400$),
the real asymptotic decay is not observed.

Here we emphasize that such a crossover at $\tau_c$ is not induced by an insufficient particle number of the lattice
in the calculation of the heat current correlation $c(\tau)$.
First, the curves with different $N$ well overlap with each other around the crossover regime.
Second, and even more important, it is clearly seen by comparing those curves that
the smaller the $N$ the faster is the $c(\tau)$ decay.
Thus the slowing down of the decay is not a consequence of an insufficient particle number $N$.

It is commonly accepted that the length dependence of the heat conductivity $\kappa_{\rm NE}(L)$ follows the $\kappa_{\rm GK}(L)$ defined below,
\begin{equation}
 \kappa_{\rm GK}(L) \equiv \int_0^{t} c(\tau) d\tau,
\label{eq:GK}
\end{equation}
where $t=L/v_s$ with at most a constant-factor difference as $L\rightarrow \infty$\cite{PhysRep.377.1,*AdvPhys.57.457}.
If $c(\tau)$ decays asymptotically with $\tau$ as $\tau^{\gamma}$
($\gamma>-1$), then the heat conductivity should
diverge with $L$ as $L^{\alpha}$ ($\alpha=1+\gamma$).
Otherwise, if $c(\tau)$ decays faster than $\tau^{-1}$,
then the heat conductivity converges.
The speed of sound, $v_s$, can be obtained by simulating the heat diffusion process
\cite{Chaos.15.015121,*EPJB.85.337}. In the FPU-$\alpha\beta$ lattices $v_s \approx 1.5$.

At first glance, since $c(\tau)$ displays a slow power-law
decay $\tau^{-2/3}$ from $\tau=\tau_c=700$ upwards, it seems reasonable to predict that $\kappa$ follows
$L^{1/3}$ from $L=L_c\equiv\tau_c v_s$ upwards.
However, the finding in Fig.~\ref{fig:fpua2T1heat} is evidently inconsistent with this expectation.
To explain this inconsistency, we suppose, as is observed in Fig.~\ref{fig:JJfpuaT1},
when $\tau$ is greater than a threshold value $\tau_c$, $c(\tau)$ follows $B\tau^{\gamma}$.
Then for $L>v_s\tau_c$, the length dependence of the heat conductivity is
\begin{align}
 \kappa_{\rm GK}(L) &\equiv \int_{0}^{L/v_s} c(\tau) d\tau =A +  B \int_{\tau_c}^{L/v_s} \tau^{\gamma} d\tau \nonumber\\
  &=        \frac{B}{1+\gamma}[(\frac{L}{v_s})^{1+\gamma} + (\frac{A(1+\gamma)}{B} - {\tau_c}^{1+\gamma})] \nonumber\\
  &=  \frac{B}{1+\gamma}[(\frac{L}{v_s})^{1+\gamma} + D],
\end{align}
where
$A \equiv \int_0^{\tau_c} c(\tau) d\tau$ and $D\equiv \frac{A(1+\gamma)}{B} - {\tau_c}^{1+\gamma}$ are constants independent of $L$.
The running slope of $\kappa_{\rm GK}(L)$ in double logarithmic scale equals
\begin{eqnarray} \label{runningslpoe}
 \alpha(L) \equiv \frac{d\ln \kappa_{\rm GK}(L)}{d\ln L}
 =\frac{(1+\gamma)(\frac{L}{v_s})^{1+\gamma}}{(\frac{L}{v_s})^{1+\gamma}+D}.
\end{eqnarray}
When $L\rightarrow v_s\tau_c+$, $\alpha(L)\rightarrow\frac BA\tau_c^{1+\gamma}$.
In the thermodynamic limit, $\alpha \rightarrow1+\gamma$.

Therefore, in order to observe a value of $\alpha \in(\frac BA\tau_c^{1+\gamma},1+\gamma)$ by the nonequilibrium heat bath method,
we need to simulate a lattice with a length of at least
\begin{eqnarray}
 L(\alpha)=v_s(\frac{D}{\frac{1+\gamma}{\alpha}-1})^{\frac1{1+\gamma}}.
\end{eqnarray}
For the FPU-$\alpha\beta$ lattice, $v_s=1.5$. According to Fig.~\ref{fig:JJfpuaT1}, $\tau_c=7\times 10^{2}$, $A=59$, $\gamma=-2/3$, and $B=0.7$.
Thus $D=19.2$ and $\alpha(L)\in (0.1,1/3)$.
To observe $\alpha$ close to its asymptotic value $1/3$, say, $\alpha=0.3$,
a simulation with length $L(0.3)=7.7\times 10^{6}$ is necessary.
In contrast, the longest lattice that has already been simulated is only two orders of magnitude shorter.
This clearly explains why a flat $\kappa$ has been observed in Fig.~\ref{fig:fpua2T1heat}.

\begin{figure}[ht]
\includegraphics[width=\columnwidth]{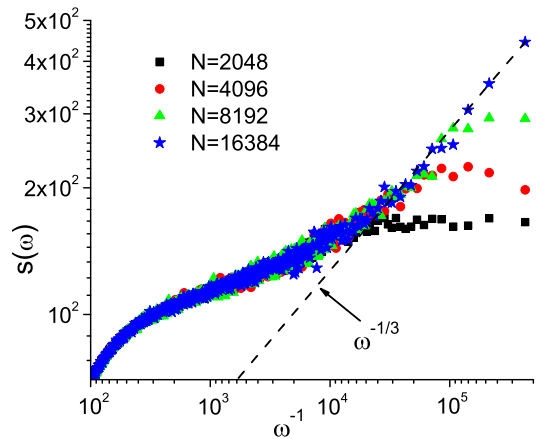}
\vspace{-1.0cm} \caption{\label{fig:fpuafftT1} (color online).
The power spectrum $S(\omega)$ of the global heat current of the FPU-$\alpha\beta$ lattices for various length $N$.
To reduce statistical fluctuations, data binning over contiguous frequencies was performed.
$S(\omega)$ for the largest $N=16384$ is from an average over 1000 sets of time series.
}
\end{figure}

The third method we use to give support to such a crossover is
the power spectrum $S(\omega)$ of the global heat current. In momentum-conserving lattices,
heat conduction in the thermodynamic limit is dominated by low-frequency
phonons. According to the Wiener-Khinchin theorem, in the low-frequency limit, the heat
current autocorrelation function $c(\tau)\sim\tau^\gamma$ corresponds to
$S(\omega)\sim\omega^{-\gamma-1}$ if $\gamma>-1$, and a flat
$S(\omega)$ otherwise~\cite{EurophysLett.43.271,*PhysRevE.68.067102}.
The power spectrum $S(\omega)$ of the FPU-$\alpha\beta$ lattices with different particle numbers are depicted in Fig.~\ref{fig:fpuafftT1}.
The conditions of the calculations are the same as those we use to calculate $c(\tau)$.
A flat $S(\omega)$ regime, $\omega \in (1\times 10^{-4} \sim 5\times 10^{-3})$ is observed
\footnote{As a comparison, there is no such a transient process in
the 1D purely quartic model (namely in Eq.~\ref{eq:fpuab} $k_2=k_3=0$ and $k_4=1$).
$S(\omega)$ directly approaches its asymptotic divergence from a much higher value of $\omega$~\cite{PhysRevLett.105.160601}.}.
This regime clearly corresponds to the flat $\kappa$ regime shown in Fig.~\ref{fig:fpua2T1heat},
as well as the fast exponential-like decay regime of $c(\tau)$ shown in Fig.~\ref{fig:JJfpuaT1}.
However, in yet a lower $\omega$ regime, $S(\omega)$ regains its slope, approaching $\omega^{-1/3}$.
This corresponds to the curving up of $\kappa$ shown in Fig.~\ref{fig:fpua2T1heat},
as well as the slow power-law decay of $c(\tau)$ shown in Fig.~\ref{fig:JJfpuaT1}.
By comparing curves for different $N$, we conclude that such a crossover is definitely {\it not} induced by a finite $N$ effect.

The conclusions by the three different methods are well consistent with each other.
All of them confirm the theoretical expectation that heat conduction generally displays a power-law divergence in
1D nonlinear lattices with total momentum conservation.
Asymmetric interparticle interaction can only induce a transient finite-size flat
$\kappa$ whereas its asymptotic divergence remains unchanged.

The temperature is quite important in this model
because when it is higher the interparticle interaction is dominated more by the symmetric quartic term.
To check the role of the temperature,
we have also calculated $c(\tau)$ in the same lattice but with a much lower energy density $\epsilon=0.0977$,
which corresponds to $T=0.1$.
Similarly, a fast exponential-like decay is observed in a range of $\tau$
until $\tau$ reaches a critical value $\tau_c \approx 2\times 10^3$, which is much longer (see the inset in Fig.\ref{fig:JJfpuaT1}).
However, the asymptotic decay remains unchanged again.

\begin{figure}[ht]
\includegraphics[width=\columnwidth]{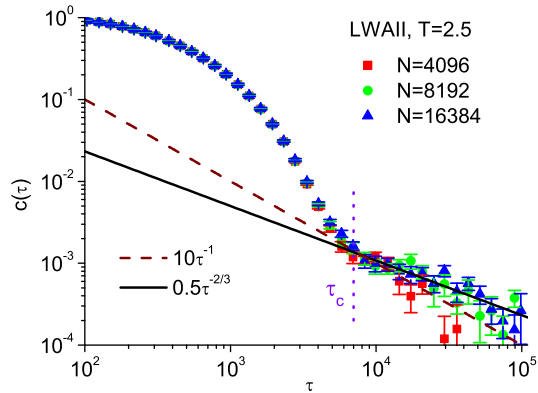}
\vspace{-1.0cm} \caption{\label{fig:JJexp} (color online).
The rescaled heat current autocorrelation function $c(\tau)$ of LWAII.
Lines with slope $-1$ and $-2/3$ are drawn for reference.
Please focus on the data for the largest $N=16384$, which cost the most computational resources.
That $c(\tau)$ decays asymptotically more slowly than $\tau^{-1}$ is highly unlikely.}
\end{figure}

Finally we study lattices with asymmetric interparticle
interactions, called the LWAII model in Ref.~\cite{PhysRevE.85.060102},
\begin{eqnarray}
 V(x)=\frac12(x+1)^2+e^{-x}.
\end{eqnarray}
Such an asymmetry induces an even stronger finite-size effect.
We have also calculated its rescaled heat current autocorrelation function $c(\tau)$ at temperature $T=2.5$,
which corresponds to $\epsilon=2.403$ (see Fig.~\ref{fig:JJexp}).
According to Ref.~\cite{PhysRevE.85.060102}, heat conduction at this temperature obeys the Fourier law.
Once more we observe an exponential-like decay in an even wider range of $\tau$.
The heat current correlation $c(\tau)$ also ceases this fast decay and changes to a slow power-law decay
as $\tau>\tau_c\approx 7\times 10^{3}$.
Calculation for each lattice was performed on one or more Nvidia
Tesla-2075 graphics processing units (GPUs), each of which has 448 Cuda processing cores on board.
This part of the calculation alone cost several months of wall time in
our eight-GPU graphics workstation.
Due to rather big fluctuations, we are still not able to obtain the asymptotic power exponent $\gamma$ with satisfactory accuracy.
However, $\gamma<-1$ is highly unlikely.

\begin{figure}[ht]
\includegraphics[width=\columnwidth]{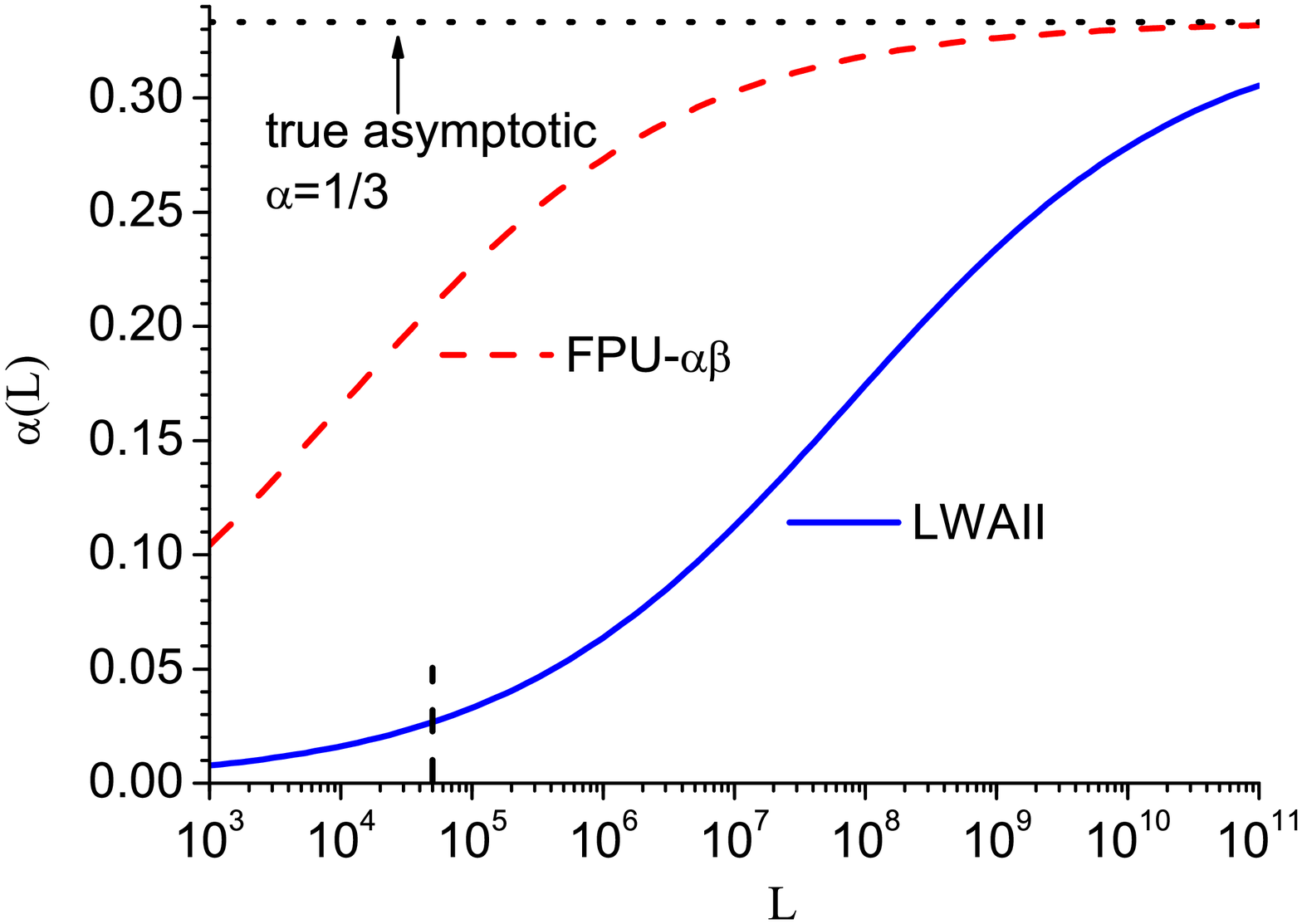}
\vspace{-1.0cm} \caption{\label{fig:avsN} (color online).
The running slope $\alpha$ of the heat conductivity as a function of the required length of lattice, $L$, in nonequilibrium heat bath method
for the FPU-$\alpha\beta$ lattice and LWAII. The longest lattice that has been
simulated for LWAII ($L=5 \times 10^4$) is indicated by the black vertical dashed line.
To observe $\alpha$ approaching 0.3, simulating a lattice which is greater than {\it six} orders of magnitude longer is necessary.
 }
\end{figure}

The theoretical expected value $\gamma=-2/3$ is again a reasonable estimate.
For the LWAII model $v_s=2.0$.
According to Fig.~\ref{fig:JJexp},
$\tau_c=7\times 10^{3}$, $A=640$, and $B=0.5$.
Thus $D=408$ and $\alpha(L)\in (0.015,1/3)$.
In Fig.~\ref{fig:avsN}, we plot the running slope $\alpha$
as a function of the required length $L$.
The longest lattice that is studied in Ref.~\cite{PhysRevE.85.060102} is $L=5 \times 10^4$
(indicated by the black vertical dashed line in the figure) where $\alpha(L)=0.022$.
The value is very close to zero.
The Fourier-like law that has been observed is thus understandable.
To observe $\alpha$ approaching $0.3$,
simulating a lattice with length $L(0.3)=1\times10^{11}$, which is more than {\it six} orders of magnitude longer, is required.
We have to say that it is unlikely to achieve such a goal in the near future.
Such a length is in the macroscopic scale already (10 m if the lattice constant is 1$\AA$).
The result for the FPU-$\alpha\beta$ model is also plotted for reference.

\section{discussion}

In summary, we studied heat conduction in a few 1D lattices with asymmetric interparticle interactions.
Strong finite-size effects are observed in all cases, which make the heat conductivity $\kappa$ obey a Fourier-like law in a wide regime of the lattice length $L$.
The heat current autocorrelation function $c(\tau)$ accordingly displays a fast exponential-like decay in a wide range of values of $\tau$.
These agree with the observations made in existing studies~\cite{PhysRevE.85.060102,2012arXiv1204.5933C} very well.
This phenomenon can be attributed to the thermal pressure, which only exists in lattices with asymmetric interparticle interactions.
The thermal pressure makes the particle densities vary with instantaneous local temperatures.
This effect induces a kind of disorder, which provides a mechanism to scatter phonons that process heat transport.
However, in momentum-conserving lattices the asymptotic behavior of heat conduction is dominated by long-wavelength (low-frequency) phonons.
The particle density fluctuation is negligible in large spatial scales.
As a consequence, in a yet longer-$\tau$ regime, $c(\tau)$ recovers a slow power-law decay with an exponent less negative than $-1$.
Accordingly the heat conductivity $\kappa$ should asymptotically recover the power-law divergence as well.
This conclusion is also consistent with some very recent studies~\cite{2013arXiv1307.4725S,*2013arXiv1308.5475D}.
However, it is very hard to directly observe this asymptotic behavior by the nonequilibrium heat bath method
because the required lengths of lattice are very long.
By analyzing the data of $c(\tau)$, we are able to predict quantitatively those necessary lengths.
Their values are so long that the computational demand is beyond the present and near-future capabilities.

\begin{acknowledgments}
This work is supported in part by the National Natural Science
Foundation of China under Grant No. 11275267 (LW),
the Program for New Century Excellent Talents in University in China under Grant No. NCET-10-0800 (LW),
and a startup fund from Tongji University (BL).
LW would like to thank C.H. Lai for useful discussions.
Computational resources were provided by the Physical Laboratory of High Performance Computing at Renmin University of China.
\end{acknowledgments}


\begin{thebibliography}{31}%
\makeatletter
\providecommand \@ifxundefined [1]{%
 \@ifx{#1\undefined}
}%
\providecommand \@ifnum [1]{%
 \ifnum #1\expandafter \@firstoftwo
 \else \expandafter \@secondoftwo
 \fi
}%
\providecommand \@ifx [1]{%
 \ifx #1\expandafter \@firstoftwo
 \else \expandafter \@secondoftwo
 \fi
}%
\providecommand \natexlab [1]{#1}%
\providecommand \enquote  [1]{``#1''}%
\providecommand \bibnamefont  [1]{#1}%
\providecommand \bibfnamefont [1]{#1}%
\providecommand \citenamefont [1]{#1}%
\providecommand \href@noop [0]{\@secondoftwo}%
\providecommand \href [0]{\begingroup \@sanitize@url \@href}%
\providecommand \@href[1]{\@@startlink{#1}\@@href}%
\providecommand \@@href[1]{\endgroup#1\@@endlink}%
\providecommand \@sanitize@url [0]{\catcode `\\12\catcode `\$12\catcode
  `\&12\catcode `\#12\catcode `\^12\catcode `\_12\catcode `\%12\relax}%
\providecommand \@@startlink[1]{}%
\providecommand \@@endlink[0]{}%
\providecommand \url  [0]{\begingroup\@sanitize@url \@url }%
\providecommand \@url [1]{\endgroup\@href {#1}{\urlprefix }}%
\providecommand \urlprefix  [0]{URL }%
\providecommand \Eprint [0]{\href }%
\providecommand \doibase [0]{http://dx.doi.org/}%
\providecommand \selectlanguage [0]{\@gobble}%
\providecommand \bibinfo  [0]{\@secondoftwo}%
\providecommand \bibfield  [0]{\@secondoftwo}%
\providecommand \translation [1]{[#1]}%
\providecommand \BibitemOpen [0]{}%
\providecommand \bibitemStop [0]{}%
\providecommand \bibitemNoStop [0]{.\EOS\space}%
\providecommand \EOS [0]{\spacefactor3000\relax}%
\providecommand \BibitemShut  [1]{\csname bibitem#1\endcsname}%
\let\auto@bib@innerbib\@empty
\bibitem [{\citenamefont {Lepri}\ \emph
  {et~al.}(2003{\natexlab{a}})\citenamefont {Lepri}, \citenamefont {Livi},\
  and\ \citenamefont {Politi}}]{PhysRep.377.1}%
  \BibitemOpen
  \bibfield  {author} {\bibinfo {author} {\bibfnamefont {S.}~\bibnamefont
  {Lepri}}, \bibinfo {author} {\bibfnamefont {R.}~\bibnamefont {Livi}}, \ and\
  \bibinfo {author} {\bibfnamefont {A.}~\bibnamefont {Politi}},\ }\href@noop {}
  {\bibfield  {journal} {\bibinfo  {journal} {Phys. Rep.}\ }\textbf {\bibinfo
  {volume} {377}},\ \bibinfo {pages} {1} (\bibinfo {year}
  {2003}{\natexlab{a}})}\BibitemShut {NoStop}%
\bibitem [{\citenamefont {Dhar}(2008)}]{AdvPhys.57.457}%
  \BibitemOpen
  \bibfield  {author} {\bibinfo {author} {\bibfnamefont {A.}~\bibnamefont
  {Dhar}},\ }\href {\doibase 10.1080/00018730802538522} {\bibfield  {journal}
  {\bibinfo  {journal} {Adv. Phys.}\ }\textbf {\bibinfo {volume} {57}},\
  \bibinfo {pages} {457} (\bibinfo {year} {2008})}\BibitemShut {NoStop}%
\bibitem [{\citenamefont {Lepri}\ \emph {et~al.}(1997)\citenamefont {Lepri},
  \citenamefont {Livi},\ and\ \citenamefont {Politi}}]{PhysRevLett.78.1896}%
  \BibitemOpen
  \bibfield  {author} {\bibinfo {author} {\bibfnamefont {S.}~\bibnamefont
  {Lepri}}, \bibinfo {author} {\bibfnamefont {R.}~\bibnamefont {Livi}}, \ and\
  \bibinfo {author} {\bibfnamefont {A.}~\bibnamefont {Politi}},\ }\href
  {\doibase 10.1103/PhysRevLett.78.1896} {\bibfield  {journal} {\bibinfo
  {journal} {Phys. Rev. Lett.}\ }\textbf {\bibinfo {volume} {78}},\ \bibinfo
  {pages} {1896} (\bibinfo {year} {1997})}\BibitemShut {NoStop}%
\bibitem [{\citenamefont {Hatano}(1999)}]{PhysRevE.59.R1}%
  \BibitemOpen
  \bibfield  {author} {\bibinfo {author} {\bibfnamefont {T.}~\bibnamefont
  {Hatano}},\ }\href {\doibase 10.1103/PhysRevE.59.R1} {\bibfield  {journal}
  {\bibinfo  {journal} {Phys. Rev. E}\ }\textbf {\bibinfo {volume} {59}},\
  \bibinfo {pages} {R1} (\bibinfo {year} {1999})}\BibitemShut {NoStop}%
\bibitem [{\citenamefont {Mai}\ \emph {et~al.}(2007)\citenamefont {Mai},
  \citenamefont {Dhar},\ and\ \citenamefont {Narayan}}]{PhysRevLett.98.184301}%
  \BibitemOpen
  \bibfield  {author} {\bibinfo {author} {\bibfnamefont {T.}~\bibnamefont
  {Mai}}, \bibinfo {author} {\bibfnamefont {A.}~\bibnamefont {Dhar}}, \ and\
  \bibinfo {author} {\bibfnamefont {O.}~\bibnamefont {Narayan}},\ }\href
  {\doibase 10.1103/PhysRevLett.98.184301} {\bibfield  {journal} {\bibinfo
  {journal} {Phys. Rev. Lett.}\ }\textbf {\bibinfo {volume} {98}},\ \bibinfo
  {pages} {184301} (\bibinfo {year} {2007})}\BibitemShut {NoStop}%
\bibitem [{\citenamefont {Giardin\'a}\ \emph {et~al.}(2000)\citenamefont
  {Giardin\'a}, \citenamefont {Livi}, \citenamefont {Politi},\ and\
  \citenamefont {Vassalli}}]{PhysRevLett.84.2144}%
  \BibitemOpen
  \bibfield  {author} {\bibinfo {author} {\bibfnamefont {C.}~\bibnamefont
  {Giardin\'a}}, \bibinfo {author} {\bibfnamefont {R.}~\bibnamefont {Livi}},
  \bibinfo {author} {\bibfnamefont {A.}~\bibnamefont {Politi}}, \ and\ \bibinfo
  {author} {\bibfnamefont {M.}~\bibnamefont {Vassalli}},\ }\href {\doibase
  10.1103/PhysRevLett.84.2144} {\bibfield  {journal} {\bibinfo  {journal}
  {Phys. Rev. Lett.}\ }\textbf {\bibinfo {volume} {84}},\ \bibinfo {pages}
  {2144} (\bibinfo {year} {2000})}\BibitemShut {NoStop}%
\bibitem [{\citenamefont {Gendelman}\ and\ \citenamefont
  {Savin}(2000)}]{PhysRevLett.84.2381}%
  \BibitemOpen
  \bibfield  {author} {\bibinfo {author} {\bibfnamefont {O.~V.}\ \bibnamefont
  {Gendelman}}\ and\ \bibinfo {author} {\bibfnamefont {A.~V.}\ \bibnamefont
  {Savin}},\ }\href {\doibase 10.1103/PhysRevLett.84.2381} {\bibfield
  {journal} {\bibinfo  {journal} {Phys. Rev. Lett.}\ }\textbf {\bibinfo
  {volume} {84}},\ \bibinfo {pages} {2381} (\bibinfo {year}
  {2000})}\BibitemShut {NoStop}%
\bibitem [{\citenamefont {Prosen}\ and\ \citenamefont
  {Campbell}(2000)}]{PhysRevLett.84.2857}%
  \BibitemOpen
  \bibfield  {author} {\bibinfo {author} {\bibfnamefont {T.}~\bibnamefont
  {Prosen}}\ and\ \bibinfo {author} {\bibfnamefont {D.~K.}\ \bibnamefont
  {Campbell}},\ }\href {\doibase 10.1103/PhysRevLett.84.2857} {\bibfield
  {journal} {\bibinfo  {journal} {Phys. Rev. Lett.}\ }\textbf {\bibinfo
  {volume} {84}},\ \bibinfo {pages} {2857} (\bibinfo {year}
  {2000})}\BibitemShut {NoStop}%
\bibitem [{\citenamefont {Narayan}\ and\ \citenamefont
  {Ramaswamy}(2002)}]{PhysRevLett.89.200601}%
  \BibitemOpen
  \bibfield  {author} {\bibinfo {author} {\bibfnamefont {O.}~\bibnamefont
  {Narayan}}\ and\ \bibinfo {author} {\bibfnamefont {S.}~\bibnamefont
  {Ramaswamy}},\ }\href {\doibase 10.1103/PhysRevLett.89.200601} {\bibfield
  {journal} {\bibinfo  {journal} {Phys. Rev. Lett.}\ }\textbf {\bibinfo
  {volume} {89}},\ \bibinfo {pages} {200601} (\bibinfo {year}
  {2002})}\BibitemShut {NoStop}%
\bibitem [{\citenamefont {Mai}\ and\ \citenamefont
  {Narayan}(2006)}]{PhysRevE.73.061202}%
  \BibitemOpen
  \bibfield  {author} {\bibinfo {author} {\bibfnamefont {T.}~\bibnamefont
  {Mai}}\ and\ \bibinfo {author} {\bibfnamefont {O.}~\bibnamefont {Narayan}},\
  }\href {\doibase 10.1103/PhysRevE.73.061202} {\bibfield  {journal} {\bibinfo
  {journal} {Phys. Rev. E}\ }\textbf {\bibinfo {volume} {73}},\ \bibinfo
  {pages} {061202} (\bibinfo {year} {2006})}\BibitemShut {NoStop}%
\bibitem [{\citenamefont {Delfini}\ \emph {et~al.}(2006)\citenamefont
  {Delfini}, \citenamefont {Lepri}, \citenamefont {Livi},\ and\ \citenamefont
  {Politi}}]{PhysRevE.73.060201}%
  \BibitemOpen
  \bibfield  {author} {\bibinfo {author} {\bibfnamefont {L.}~\bibnamefont
  {Delfini}}, \bibinfo {author} {\bibfnamefont {S.}~\bibnamefont {Lepri}},
  \bibinfo {author} {\bibfnamefont {R.}~\bibnamefont {Livi}}, \ and\ \bibinfo
  {author} {\bibfnamefont {A.}~\bibnamefont {Politi}},\ }\href {\doibase
  10.1103/PhysRevE.73.060201} {\bibfield  {journal} {\bibinfo  {journal} {Phys.
  Rev. E}\ }\textbf {\bibinfo {volume} {73}},\ \bibinfo {pages} {060201}
  (\bibinfo {year} {2006})}\BibitemShut {NoStop}%
\bibitem [{\citenamefont {Lepri}\ \emph {et~al.}(1998)\citenamefont {Lepri},
  \citenamefont {Livi},\ and\ \citenamefont {Politi}}]{EurophysLett.43.271}%
  \BibitemOpen
  \bibfield  {author} {\bibinfo {author} {\bibfnamefont {S.}~\bibnamefont
  {Lepri}}, \bibinfo {author} {\bibfnamefont {R.}~\bibnamefont {Livi}}, \ and\
  \bibinfo {author} {\bibfnamefont {A.}~\bibnamefont {Politi}},\ }\href@noop {}
  {\bibfield  {journal} {\bibinfo  {journal} {Europhys. Lett.}\ }\textbf
  {\bibinfo {volume} {43}},\ \bibinfo {pages} {271} (\bibinfo {year}
  {1998})}\BibitemShut {NoStop}%
\bibitem [{\citenamefont {Lepri}\ \emph
  {et~al.}(2003{\natexlab{b}})\citenamefont {Lepri}, \citenamefont {Livi},\
  and\ \citenamefont {Politi}}]{PhysRevE.68.067102}%
  \BibitemOpen
  \bibfield  {author} {\bibinfo {author} {\bibfnamefont {S.}~\bibnamefont
  {Lepri}}, \bibinfo {author} {\bibfnamefont {R.}~\bibnamefont {Livi}}, \ and\
  \bibinfo {author} {\bibfnamefont {A.}~\bibnamefont {Politi}},\ }\href
  {\doibase 10.1103/PhysRevE.68.067102} {\bibfield  {journal} {\bibinfo
  {journal} {Phys. Rev. E}\ }\textbf {\bibinfo {volume} {68}},\ \bibinfo
  {pages} {067102} (\bibinfo {year} {2003}{\natexlab{b}})}\BibitemShut
  {NoStop}%
\bibitem [{\citenamefont {Basile}\ \emph {et~al.}(2006)\citenamefont {Basile},
  \citenamefont {Bernardin},\ and\ \citenamefont
  {Olla}}]{PhysRevLett.96.204303}%
  \BibitemOpen
  \bibfield  {author} {\bibinfo {author} {\bibfnamefont {G.}~\bibnamefont
  {Basile}}, \bibinfo {author} {\bibfnamefont {C.}~\bibnamefont {Bernardin}}, \
  and\ \bibinfo {author} {\bibfnamefont {S.}~\bibnamefont {Olla}},\ }\href
  {\doibase 10.1103/PhysRevLett.96.204303} {\bibfield  {journal} {\bibinfo
  {journal} {Phys. Rev. Lett.}\ }\textbf {\bibinfo {volume} {96}},\ \bibinfo
  {pages} {204303} (\bibinfo {year} {2006})}\BibitemShut {NoStop}%
\bibitem [{\citenamefont {van Beijeren}(2012)}]{PhysRevLett.108.180601}%
  \BibitemOpen
  \bibfield  {author} {\bibinfo {author} {\bibfnamefont {H.}~\bibnamefont {van
  Beijeren}},\ }\href {\doibase 10.1103/PhysRevLett.108.180601} {\bibfield
  {journal} {\bibinfo  {journal} {Phys. Rev. Lett.}\ }\textbf {\bibinfo
  {volume} {108}},\ \bibinfo {pages} {180601} (\bibinfo {year}
  {2012})}\BibitemShut {NoStop}%
\bibitem [{\citenamefont {Chang}\ \emph {et~al.}(2008)\citenamefont {Chang},
  \citenamefont {Okawa}, \citenamefont {Garcia}, \citenamefont {Majumdar},\
  and\ \citenamefont {Zettl}}]{PhysRevLett.101.075903}%
  \BibitemOpen
  \bibfield  {author} {\bibinfo {author} {\bibfnamefont {C.~W.}\ \bibnamefont
  {Chang}}, \bibinfo {author} {\bibfnamefont {D.}~\bibnamefont {Okawa}},
  \bibinfo {author} {\bibfnamefont {H.}~\bibnamefont {Garcia}}, \bibinfo
  {author} {\bibfnamefont {A.}~\bibnamefont {Majumdar}}, \ and\ \bibinfo
  {author} {\bibfnamefont {A.}~\bibnamefont {Zettl}},\ }\href {\doibase
  10.1103/PhysRevLett.101.075903} {\bibfield  {journal} {\bibinfo  {journal}
  {Phys. Rev. Lett.}\ }\textbf {\bibinfo {volume} {101}},\ \bibinfo {pages}
  {075903} (\bibinfo {year} {2008})}\BibitemShut {NoStop}%
\bibitem [{\citenamefont {Ghosh}\ \emph {et~al.}(2010)\citenamefont {Ghosh},
  \citenamefont {Bao}, \citenamefont {Nika}, \citenamefont {Subrina},
  \citenamefont {Pokatilov}, \citenamefont {Lau},\ and\ \citenamefont
  {Balandin}}]{NatMater.9.555}%
  \BibitemOpen
  \bibfield  {author} {\bibinfo {author} {\bibfnamefont {S.}~\bibnamefont
  {Ghosh}}, \bibinfo {author} {\bibfnamefont {W.}~\bibnamefont {Bao}}, \bibinfo
  {author} {\bibfnamefont {D.~L.}\ \bibnamefont {Nika}}, \bibinfo {author}
  {\bibfnamefont {S.}~\bibnamefont {Subrina}}, \bibinfo {author} {\bibfnamefont
  {E.~P.}\ \bibnamefont {Pokatilov}}, \bibinfo {author} {\bibfnamefont {C.~N.}\
  \bibnamefont {Lau}}, \ and\ \bibinfo {author} {\bibfnamefont {A.~A.}\
  \bibnamefont {Balandin}},\ }\href@noop {} {\bibfield  {journal} {\bibinfo
  {journal} {Nat. Mater.}\ }\textbf {\bibinfo {volume} {9}},\ \bibinfo {pages}
  {555} (\bibinfo {year} {2010})}\BibitemShut {NoStop}%
\bibitem [{\citenamefont {Wang}\ \emph {et~al.}(2011)\citenamefont {Wang},
  \citenamefont {Xie}, \citenamefont {Bui}, \citenamefont {Liu}, \citenamefont
  {Ni}, \citenamefont {Li},\ and\ \citenamefont {Thong}}]{NanoLett.11.113}%
  \BibitemOpen
  \bibfield  {author} {\bibinfo {author} {\bibfnamefont {Z.}~\bibnamefont
  {Wang}}, \bibinfo {author} {\bibfnamefont {R.}~\bibnamefont {Xie}}, \bibinfo
  {author} {\bibfnamefont {C.~T.}\ \bibnamefont {Bui}}, \bibinfo {author}
  {\bibfnamefont {D.}~\bibnamefont {Liu}}, \bibinfo {author} {\bibfnamefont
  {X.}~\bibnamefont {Ni}}, \bibinfo {author} {\bibfnamefont {B.}~\bibnamefont
  {Li}}, \ and\ \bibinfo {author} {\bibfnamefont {J.~T.~L.}\ \bibnamefont
  {Thong}},\ }\href@noop {} {\bibfield  {journal} {\bibinfo  {journal} {Nano
  Lett.}\ }\textbf {\bibinfo {volume} {11}},\ \bibinfo {pages} {113} (\bibinfo
  {year} {2011})}\BibitemShut {NoStop}%
\bibitem [{\citenamefont {Nika}\ \emph {et~al.}(2012)\citenamefont {Nika},
  \citenamefont {Askerov},\ and\ \citenamefont {Balandin}}]{NanoLett.12.3238}%
  \BibitemOpen
  \bibfield  {author} {\bibinfo {author} {\bibfnamefont {D.~L.}\ \bibnamefont
  {Nika}}, \bibinfo {author} {\bibfnamefont {A.~S.}\ \bibnamefont {Askerov}}, \
  and\ \bibinfo {author} {\bibfnamefont {A.~A.}\ \bibnamefont {Balandin}},\
  }\href@noop {} {\bibfield  {journal} {\bibinfo  {journal} {Nano. Lett.}\
  }\textbf {\bibinfo {volume} {12}},\ \bibinfo {pages} {3238} (\bibinfo {year}
  {2012})}\BibitemShut {NoStop}%
\bibitem [{\citenamefont {Li}\ \emph {et~al.}(2012)\citenamefont {Li},
  \citenamefont {Ren}, \citenamefont {Wang}, \citenamefont {Zhang},
  \citenamefont {H\"anggi},\ and\ \citenamefont {Li}}]{RevModPhys.84.1045}%
  \BibitemOpen
  \bibfield  {author} {\bibinfo {author} {\bibfnamefont {N.}~\bibnamefont
  {Li}}, \bibinfo {author} {\bibfnamefont {J.}~\bibnamefont {Ren}}, \bibinfo
  {author} {\bibfnamefont {L.}~\bibnamefont {Wang}}, \bibinfo {author}
  {\bibfnamefont {G.}~\bibnamefont {Zhang}}, \bibinfo {author} {\bibfnamefont
  {P.}~\bibnamefont {H\"anggi}}, \ and\ \bibinfo {author} {\bibfnamefont
  {B.}~\bibnamefont {Li}},\ }\href {\doibase 10.1103/RevModPhys.84.1045}
  {\bibfield  {journal} {\bibinfo  {journal} {Rev. Mod. Phys.}\ }\textbf
  {\bibinfo {volume} {84}},\ \bibinfo {pages} {1045} (\bibinfo {year}
  {2012})}\BibitemShut {NoStop}%
\bibitem [{\citenamefont {Wang}\ and\ \citenamefont
  {Wang}(2011)}]{EurophysLett.93.54002}%
  \BibitemOpen
  \bibfield  {author} {\bibinfo {author} {\bibfnamefont {L.}~\bibnamefont
  {Wang}}\ and\ \bibinfo {author} {\bibfnamefont {T.}~\bibnamefont {Wang}},\
  }\href {http://iopscience.iop.org/0295-5075/93/5/54002} {\bibfield  {journal}
  {\bibinfo  {journal} {Europhys. Lett.}\ }\textbf {\bibinfo {volume} {93}},\
  \bibinfo {pages} {54002} (\bibinfo {year} {2011})}\BibitemShut {NoStop}%
\bibitem [{\citenamefont {Kubo}\ \emph {et~al.}(1991)\citenamefont {Kubo},
  \citenamefont {Toda},\ and\ \citenamefont {Hashitsume}}]{GREENKUBO}%
  \BibitemOpen
  \bibfield  {author} {\bibinfo {author} {\bibfnamefont {R.}~\bibnamefont
  {Kubo}}, \bibinfo {author} {\bibfnamefont {M.}~\bibnamefont {Toda}}, \ and\
  \bibinfo {author} {\bibfnamefont {N.}~\bibnamefont {Hashitsume}},\
  }\href@noop {} {\emph {\bibinfo {title} {Statistical Physics II, Springer
  Series in Solid State Sciences Vol. 31}}}\ (\bibinfo  {publisher} {Springer,
  Berlin},\ \bibinfo {year} {1991})\BibitemShut {NoStop}%
\bibitem [{\citenamefont {Wang}\ \emph {et~al.}(2010)\citenamefont {Wang},
  \citenamefont {He},\ and\ \citenamefont {Hu}}]{PhysRevLett.105.160601}%
  \BibitemOpen
  \bibfield  {author} {\bibinfo {author} {\bibfnamefont {L.}~\bibnamefont
  {Wang}}, \bibinfo {author} {\bibfnamefont {D.}~\bibnamefont {He}}, \ and\
  \bibinfo {author} {\bibfnamefont {B.}~\bibnamefont {Hu}},\ }\href {\doibase
  10.1103/PhysRevLett.105.160601} {\bibfield  {journal} {\bibinfo  {journal}
  {Phys. Rev. Lett.}\ }\textbf {\bibinfo {volume} {105}},\ \bibinfo {pages}
  {160601} (\bibinfo {year} {2010})}\BibitemShut {NoStop}%
\bibitem [{\citenamefont {Wang}\ \emph {et~al.}(2012)\citenamefont {Wang},
  \citenamefont {Hu},\ and\ \citenamefont {Li}}]{PhysRevE.86.040101}%
  \BibitemOpen
  \bibfield  {author} {\bibinfo {author} {\bibfnamefont {L.}~\bibnamefont
  {Wang}}, \bibinfo {author} {\bibfnamefont {B.}~\bibnamefont {Hu}}, \ and\
  \bibinfo {author} {\bibfnamefont {B.}~\bibnamefont {Li}},\ }\href {\doibase
  10.1103/PhysRevE.86.040101} {\bibfield  {journal} {\bibinfo  {journal} {Phys.
  Rev. E}\ }\textbf {\bibinfo {volume} {86}},\ \bibinfo {pages} {040101}
  (\bibinfo {year} {2012})}\BibitemShut {NoStop}%
\bibitem [{\citenamefont {{Chen}}\ \emph {et~al.}(2012)\citenamefont {{Chen}},
  \citenamefont {{Zhang}}, \citenamefont {{Wang}},\ and\ \citenamefont
  {{Zhao}}}]{2012arXiv1204.5933C}%
  \BibitemOpen
  \bibfield  {author} {\bibinfo {author} {\bibfnamefont {S.}~\bibnamefont
  {{Chen}}}, \bibinfo {author} {\bibfnamefont {Y.}~\bibnamefont {{Zhang}}},
  \bibinfo {author} {\bibfnamefont {J.}~\bibnamefont {{Wang}}}, \ and\ \bibinfo
  {author} {\bibfnamefont {H.}~\bibnamefont {{Zhao}}},\ }\href@noop {}
  {\bibfield  {journal} {\bibinfo  {journal} {ArXiv e-prints}\ } (\bibinfo
  {year} {2012})},\ \Eprint {http://arxiv.org/abs/1204.5933} {arXiv:1204.5933
  [cond-mat.stat-mech]} \BibitemShut {NoStop}%
\bibitem [{\citenamefont {Li}\ \emph {et~al.}(2005)\citenamefont {Li},
  \citenamefont {Wang}, \citenamefont {Wang},\ and\ \citenamefont
  {Zhang}}]{Chaos.15.015121}%
  \BibitemOpen
  \bibfield  {author} {\bibinfo {author} {\bibfnamefont {B.}~\bibnamefont
  {Li}}, \bibinfo {author} {\bibfnamefont {J.}~\bibnamefont {Wang}}, \bibinfo
  {author} {\bibfnamefont {L.}~\bibnamefont {Wang}}, \ and\ \bibinfo {author}
  {\bibfnamefont {G.}~\bibnamefont {Zhang}},\ }\href@noop {} {\bibfield
  {journal} {\bibinfo  {journal} {Chaos}\ }\textbf {\bibinfo {volume} {15}},\
  \bibinfo {pages} {015121} (\bibinfo {year} {2005})}\BibitemShut {NoStop}%
\bibitem [{\citenamefont {Liu}\ \emph {et~al.}(2012)\citenamefont {Liu},
  \citenamefont {Xu}, \citenamefont {Xie}, \citenamefont {Zhang},\ and\
  \citenamefont {Li}}]{EPJB.85.337}%
  \BibitemOpen
  \bibfield  {author} {\bibinfo {author} {\bibfnamefont {S.}~\bibnamefont
  {Liu}}, \bibinfo {author} {\bibfnamefont {X.}~\bibnamefont {Xu}}, \bibinfo
  {author} {\bibfnamefont {R.}~\bibnamefont {Xie}}, \bibinfo {author}
  {\bibfnamefont {G.}~\bibnamefont {Zhang}}, \ and\ \bibinfo {author}
  {\bibfnamefont {B.}~\bibnamefont {Li}},\ }\href@noop {} {\bibfield  {journal}
  {\bibinfo  {journal} {Eur. Phys. J. B}\ }\textbf {\bibinfo {volume} {85}},\
  \bibinfo {pages} {337} (\bibinfo {year} {2012})}\BibitemShut {NoStop}%
\bibitem [{Note1()}]{Note1}%
  \BibitemOpen
  \bibinfo {note} {As a comparison, there is no such a transient process in the
  1D purely quartic model (namely in Eq.~\ref {eq:fpuab} $k_2=k_3=0$ and
  $k_4=1$). $S(\omega )$ directly approaches its asymptotic divergence from a
  much higher value of $\omega $~\cite {PhysRevLett.105.160601}.}\BibitemShut
  {Stop}%
\bibitem [{\citenamefont {Zhong}\ \emph {et~al.}(2012)\citenamefont {Zhong},
  \citenamefont {Zhang}, \citenamefont {Wang},\ and\ \citenamefont
  {Zhao}}]{PhysRevE.85.060102}%
  \BibitemOpen
  \bibfield  {author} {\bibinfo {author} {\bibfnamefont {Y.}~\bibnamefont
  {Zhong}}, \bibinfo {author} {\bibfnamefont {Y.}~\bibnamefont {Zhang}},
  \bibinfo {author} {\bibfnamefont {J.}~\bibnamefont {Wang}}, \ and\ \bibinfo
  {author} {\bibfnamefont {H.}~\bibnamefont {Zhao}},\ }\href {\doibase
  10.1103/PhysRevE.85.060102} {\bibfield  {journal} {\bibinfo  {journal} {Phys.
  Rev. E}\ }\textbf {\bibinfo {volume} {85}},\ \bibinfo {pages} {060102}
  (\bibinfo {year} {2012})}\BibitemShut {NoStop}%
\bibitem [{\citenamefont {{Savin}}\ and\ \citenamefont
  {{Kosevich}}(2013)}]{2013arXiv1307.4725S}%
  \BibitemOpen
  \bibfield  {author} {\bibinfo {author} {\bibfnamefont {A.~V.}\ \bibnamefont
  {{Savin}}}\ and\ \bibinfo {author} {\bibfnamefont {Y.~A.}\ \bibnamefont
  {{Kosevich}}},\ }\href@noop {} {\bibfield  {journal} {\bibinfo  {journal}
  {ArXiv e-prints}\ } (\bibinfo {year} {2013})},\ \Eprint
  {http://arxiv.org/abs/1307.4725} {arXiv:1307.4725 [cond-mat.stat-mech]}
  \BibitemShut {NoStop}%
\bibitem [{\citenamefont {{Das}}\ \emph {et~al.}(2013)\citenamefont {{Das}},
  \citenamefont {{Dhar}},\ and\ \citenamefont
  {{Narayan}}}]{2013arXiv1308.5475D}%
  \BibitemOpen
  \bibfield  {author} {\bibinfo {author} {\bibfnamefont {S.~G.}\ \bibnamefont
  {{Das}}}, \bibinfo {author} {\bibfnamefont {A.}~\bibnamefont {{Dhar}}}, \
  and\ \bibinfo {author} {\bibfnamefont {O.}~\bibnamefont {{Narayan}}},\
  }\href@noop {} {\bibfield  {journal} {\bibinfo  {journal} {ArXiv e-prints}\ }
  (\bibinfo {year} {2013})},\ \Eprint {http://arxiv.org/abs/1308.5475}
  {arXiv:1308.5475 [cond-mat.stat-mech]} \BibitemShut {NoStop}%
\end{thebibliography}
\end{document}